\documentclass[referee]{aa}
\usepackage{natbib,psfig,graphicx}
\def\mathnew{\mathsurround=0pt}
\def\simov#1#2{\lower .5pt\vbox{\baselineskip0pt \lineskip-.5pt
\ialign{$\mathnew#1\hfil##\hfil$\crcr#2\crcr\sim\crcr}}}

\def\MeV{Me\kern-0.11em V}
\def\keV{ke\kern-0.11em V}

\begin{document}

\title{Photometric redshifts as a tool to study the Coma cluster galaxy
populations.
\thanks{Based on observations obtained with MegaPrime/MegaCam, a joint project
of CFHT and CEA/DAPNIA, at the Canada-France-Hawaii Telescope (CFHT) which
is operated by the National Research Council (NRC) of Canada, the Institut
National des Sciences de l'Univers of the Centre National de la Recherche
Scientifique (CNRS) of France, and the University of Hawaii. This work is
 also partly based on data products produced at TERAPIX and the Canadian
Astronomy Data Centre as part of the Canada-France-Hawaii Telescope Legacy
Survey, a collaborative project of NRC and CNRS.
Based on observations obtained with MegaPrime/MegaCam, a joint project of CFHT
and CEA/DAPNIA, at the Canada-France-Hawaii Telescope (CFHT) which is operated
by the National Research Council (NRC) of Canada, the Institute National des
Sciences de l'Univers of the Centre National de la Recherche Scientifique of
France, and the University of Hawaii. Also based on data from W. M. Keck
Observatory which is operated as a scientific partnership between the
California Institute of Technology, the University of California, and NASA. It
was made possible by the generous financial support of the W. M. Keck
Foundation. }}

\author{C. Adami\inst{1} \and 
O. Ilbert\inst{1,2} \and 
R. Pell\'o\inst{3} \and 
J.C. Cuillandre\inst{4} \and
F. Durret\inst{5} \and 
A.~Mazure\inst{1} \and 
J.P.~Picat\inst{3} \and
M.P.~Ulmer\inst{1,6} 
}

\offprints{C. Adami \email{christophe.adami@oamp.fr}}

\institute{
LAM, P\^ole de l'Etoile, Site de Ch\^ateau Gombert, 38 rue Fr\'ed\'eric Joliot-Curie,
13388 Marseille 13 Cedex, France
\and
Institute for Astronomy, 2680 Woodlawn Dr., University of Hawaii, Honolulu, Hawaii 96822, USA
\and
Laboratoire d'Astrophysique de Toulouse-Tarbes, Universit\'e de Toulouse,
CNRS, 14 Av. Edouard Belin, 
31400 Toulouse, France
\and
Canada-France-Hawaii Telescope Corporation, Kamuela, HI 96743
\and
Institut d'Astrophysique de Paris, CNRS, UMR~7095, Universit\'e Pierre et 
Marie Curie, 98bis Bd Arago, 75014 Paris, France 
\and
Department Physics $\&$ Astronomy, Northwestern University, Evanston, IL 60208-2900, USA
}

\date{Accepted . Received ; Draft printed: \today}

\authorrunning{Adami et al.}

\titlerunning{Coma cluster galaxy populations}

\abstract 
{}
{We investigate the Coma cluster galaxy luminosity function (GLF) at
faint magnitudes, in particular in the u* band where basically no
studies are presently available at these magnitudes, by applying 
photometric redshift techniques. }
{ Cluster members were selected based on Probability Distribution Function from
  photometric redshift calculations applied to
deep u*, B, V, R, I images covering a region of almost 1~deg$^2$
(completeness limit R$\sim$24). In the area covered only by the u*
image, the GLF was also derived after applying a statistical background
subtraction. }
{Global and local GLFs in the B, V, R and I bands obtained with
photometric redshift selection are consistent with our previous results based
on a statistical background subtraction. 

The GLF in the u* band shows an increase of the faint end slope towards the
outer regions of the cluster. 

The analysis of the
multicolor type spatial distribution reveals that late type galaxies are
distributed in clumps in the cluster outskirts, where X-ray
substructures are also detected and where the GLF
in the u* band is steeper.}  
{We can reproduce the GLFs computed with classical 
statistical subtraction methods by applying a photometric redshift
technique. The $u*$ GLF slope is steeper in the cluster outskirts,
varying from $\alpha \sim -1$ in the cluster center to $\alpha \sim -2$
in the cluster periphery. The concentrations of faint late type galaxies
in the cluster outskirts could explain these very steep slopes, assuming
a short burst of star formation in these galaxies when entering the
cluster.}


\keywords{galaxies: clusters: individual (Coma) ;
galaxy luminosity functions}

\maketitle

\section{Introduction}\label{sec:intro}

Ever growing optical imaging surveys of galaxies for which complete
spectroscopic follow-up is impossible due to telescope limitations has
triggered the development of photometric redshift techniques
(e.g. Bolzonella et al. 2000 or Ilbert et al. 2006a and references therein).
Based on the comparison of multi-band photometry with synthetic spectral
templates, this technique can be viewed as very low resolution
spectroscopy.  The quality of photometric redshifts depends on the
wavelength range covered by the photometric survey. High-quality
photometric redshifts and related quantities require a complete and
contiguous coverage of the spectral regions of interest, in particular
around the strong spectral features, such as the $4000$\,\AA\ break or
the Lyman $\alpha$ break. This technique has proven to be a very
valuable tool for several cosmological purposes, such as deriving field
galaxy correlation functions or galaxy luminosity functions (see
e.g. Ilbert et al. 2006b, Meneux et al. 2006). It has also been applied
in the study of distant clusters of galaxies (see e.g. White et al. 2005
and references therein).

In the present paper, we apply the photometric redshift technique
to study the galaxy population of the rich Coma
cluster. Due to its proximity (z$\sim$0.023), Coma covers a large
extent over the sky (of the order of 1 deg$^2$), requiring 
a wide field camera; data in the U band are needed to determine with
a sufficient accuracy the probability that a galaxy
belongs or not to the cluster, based in particular on the 4000~\AA\
break.

As a continuation of our Coma photometric survey (e.g. Adami et
al. 2006a) which already includes deep (R$\sim$24) wide field
(42$\times$52 arcmin$^2$) B, V, R and I images, we recently acquired a
Megacam u* band image of comparable depth and field of view. In addition
to our spectroscopic redshift catalog, this allows us to compute
photometric redshifts and Probability Distribution Function (PDF
hereafter) along the Coma cluster line of sight down to the dwarf
galaxy regime.   One of the main goals of the paper is to qualify
this photometric redshift technique applied to the galaxy luminosity
function (GLF hereafter) determination. For this, we will compare the
present photometric redshift technique with the galaxy luminosity
functions computed with the same data set but with a statistical
background removal (Adami et al. 2007a and b).

We will also compute luminosity functions in the u* band based on
statistical background subtraction in the area where only the u* band
is available.

We describe our new photometric data and photometric redshift and
PDF estimates in Section 2. We compute in Section 3 the Coma GLF using the
PDF and in Section 4 the Coma u* GLF applying statistical
subtraction. The spatial distributions of multicolor galaxy types are discussed in
Section 5. Conclusions are drawn in Section 6.

In this paper we assume H$_0$ = 70 km s$^{-1}$ Mpc$^{-1}$, $\Omega _m$=0.3,
$\Omega _{\Lambda}$=0.7,  a distance to Coma of 100 Mpc, a distance modulus =
35.00, and a scale of 0.46 kpc arcsec$^{-1}$.

\section{Photometric data and photometric redshifts}

\subsection{B, V, R and I band imaging data}


These data are fully described in Adami et al. (2006a) and we give
here only the salient points. Coma was observed with a 2 field mosaic
of the CFH12K camera in 4 bands (B, V, R and I). Altogether, our data
cover a 52$\times$42~arcmin$^2$ field over the sky (the CFH12K
f.o.v. in the following) centered on the two dominant cluster galaxies
(NGC~4874 and NGC~4889) with a completeness level in R close to
R$\sim$24. The seeing conditions were all close to 1 arcsec. These data are
available at http://cencosw.oamp.fr. We have already published several
papers based on these data (Adami et al. 2005a and b, 2006a and b, and
2007a and b).

\subsection{New u$^*$ band imaging data}

New u$^*$ band data (see Fig.~\ref{fig:imageu}) including the previous
field were obtained between 2006 and 2007 with the CFH Megacam
camera. Megacam is a mosaic of
36 individual CCDs, giving a field of view (f.o.v.) of 1 deg$^2$.  The average
seeing was 1.1~arcsec. The total
exposure time was 9.66~hours, obtained by combining 58 individual spatially
dithered exposures of 10 min each.  
We show in Fig.~\ref{fig:exp} the weight map that was generated during
the reduction procedure. This gives a good idea of the exposure time
variations across the field of view (the darker the color, the shorter
the exposure time). The sigma of the distribution of the exposure times
per pixel is $\sim$30$\%$ of the mean value. However, this affects only
the gaps between individual CCDs, that represent less than 10$\%$ of the
total area.

\begin{figure}
\centering
\caption[]{Megacam u$^*$ band weight image (the darker the color, the shorter the exposure time). 
The coordinates in degrees are indicated.}
\label{fig:exp}
\end{figure}

The data reduction was performed with the Terapix tools
(http://terapix.iap.fr/) and the standard procedure applied for example to
the CFHTLS fields (McCracken et al. 2008). A $\sim$1~deg$^2$ image was
produced with a zero point set to 30.0. We extracted an object catalog
from this image with the SExtractor package (Bertin $\&$ Arnouts 1996)
with a detection threshold of 0.4 and a minimum number of pixels (above
this threshold) of~3. We derived total Kron AB magnitudes
({\bf Kron 1980}) to compute photometric redshifts.  We
corrected each detected object for galactic extinction using the
Schlegel maps. This catalog was then cross-correlated with the B, V, R,
and I catalog already available. The cross-correlation was made by
applying a technique similar to that of Adami et al. (2006a) and
optimised to take into account small astrometric differences between
CFH12K and Megacam data. We defined small regions 2 arcmin wide in which
we iteratively adjusted the galaxy positions (assuming an initial 3
arcsec identification distance) in order to find the best
identification. We could also have applied SExtractor in double-image
mode, but this would have required to degrade all the Megacam images to
the CFH12K characteristics.

We note that the seeing values slightly differed from one band to
another, and this could produce offsets in the estimated magnitude
values. We did not attempt to correct for this seeing effect before computing
the photometric redshifts. However, it was taken into account when
fitting the observed magnitudes to the synthethic SEDs by the
photometric redshift codes for objects of our spectroscopic catalog (see
following). These corrections proved to be small.

\subsection{Zero point variations across the field of view}

\begin{figure}
\centering
\mbox{\psfig{figure=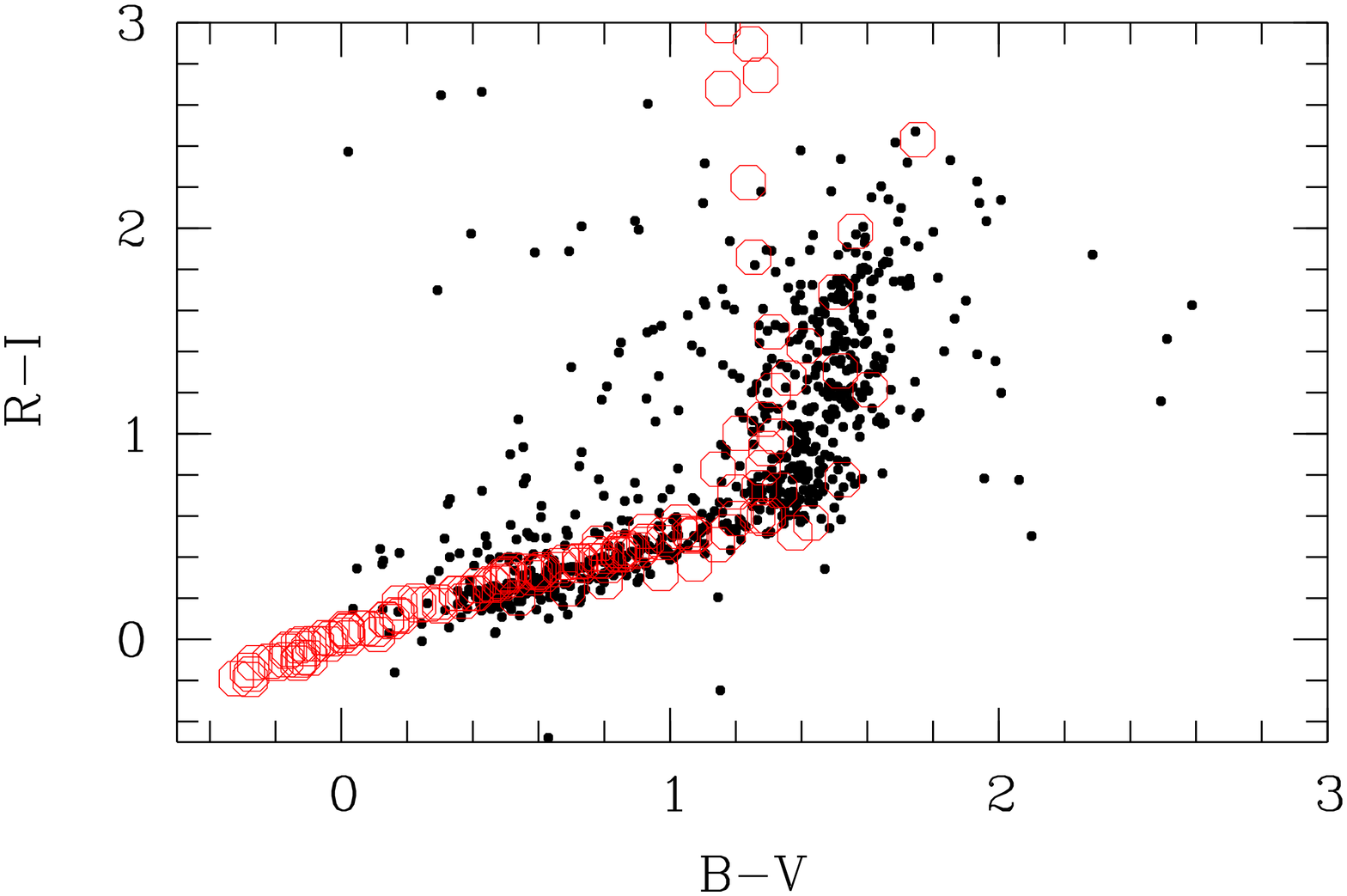,width=9cm,angle=270}}
\mbox{\psfig{figure=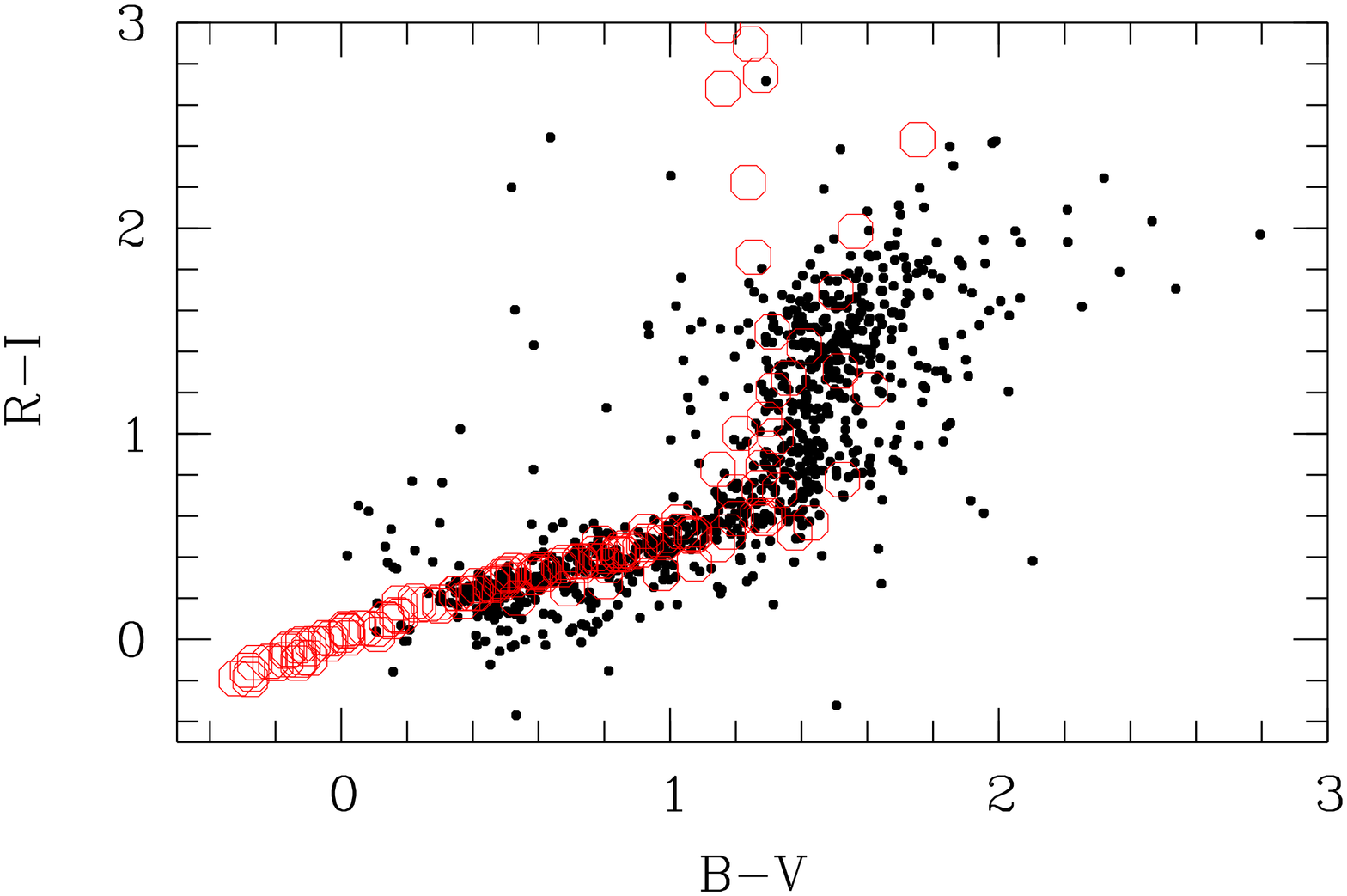,width=9cm,angle=270}}
\caption[]{R-I versus B-V for stars (as defined in Adami et
al. (2006a). Dots: observed stars, open red (grey in black and white
version) circles: empirical stellar library of Pickles (1998). The upper
and lower figures correspond to the north and south CFH12K fields
respectively.}  \label{fig:stars}
\end{figure}

Besides the possible seeing effects, it is important to assess 
possible magnitude zero point
variations across our field of view.  Given the u* data treatment, the
u* zero points are stable over the whole field of view at better than
0.05 mag (a calibration at better than $\sim$4$\%$ is achieved by the
Terapix $Elixir$ code from the field center to the field
edges). Regarding the CFH12K data, we already estimated in Adami et
al. (2006a) that correction in magnitude due to small airmass variations
would have been, at maximum, of 0.033 mag for R, 0.020 mag for V, and
0.015 mag for B, much less than the estimated errors on the
magnitudes. In the same paper (Fig. 16) we also considered
stellar tracks to estimate the additional shifts that must be applied
to the zero
points to match the empirical stellar library of Pickles (1998). The
mean (over the whole f.o.v.) corrections were negligible. However, this
does not exclude some possible small zero point variations accross the field
of view that could remain undetected due to the width of the observed stellar
tracks (of the order of 0.2 magnitude). In order to test this point, we
redid the same figure as figure~16 of Adami et al. (2006a), separating
the northern and the southern area (as defined in Adami et
al. 2007a). Fig.~\ref{fig:stars} shows that we nearly have the same
agreement between the empirical stellar library of Pickles (1998) and
the observed stellar tracks for both the northern and southern
fields. We estimated the mean magnitude shift between the northern and
southern fields to be smaller than 0.05 magnitude.

\subsection{Completeness of the u$^*$ image}

The good efficiency of Megacam in the blue allowed to reach a
similar depth in u$^*$ as in the B ($\sim$24.75 for point like objects at the
90$\%$ level, see Adami et al. 2006a), V ($\sim$24), R ($\sim$24), and I
($\sim$23.25) images.  These new data were required to compute
photometric redshifts in order to have a blue photometric band,
encompassing the 4000~\AA\ break.

The percentage of R detections with an associated u* detection is
between 87\% and 92$\%$.  This leads to estimate the 90$\%$ completeness
level of the u* image to R$\sim$24.  The $\sim$10$\%$ of galaxies
detected in R and not in u* are, at least partially, objects that are
not forming stars and which therefore do not appear in u* even in a deep
exposure.

\begin{figure*}
\caption[]{Megacam u$^*$ band image overplotted with the 5 detected
groups (circles 
correspond to 300 kpc in diameter at z=0.1) and the Neumann et
al. (2003) X-ray substructures. We corrected for a small overall shift
between the XMM and the Megacam astrometry.}  \label{fig:imageu}
\end{figure*}

\begin{figure}
\centering \mbox{\psfig{figure=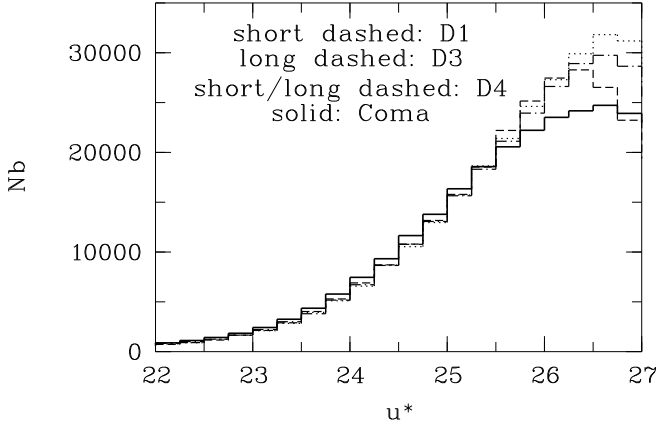,width=9cm,angle=270}}
\caption[]{Coma line of sight u* counts (solid line) compared to three
deeper u* fields from the CFHTLS: D1 (short dashed line), D3 (long
dashed line), and D4 (short/long dashed line).}  
\label{fig:ccount}
\end{figure}

In order to assess the completeness level of our u* image in another way, 
we directly compared our u* counts with the u* counts of the CFHTLS D1, D3 and D4 fields
(T0004 release). These
fields have long enough exposure times (see http://terapix.iap.fr/cplt/table-syn-T0004.html) to be
comparable with our data. We did not consider the D2 field that only has a total 
exposure time of 1.3 hours in u*. Fig.~\ref{fig:ccount} shows that the Coma counts dominate the D1,
D3, and D4 counts for u*$\leq$25, due to the presence of the highly
populated Coma cluster in the field. Our data become incomplete at
fainter magnitudes, leading to estimate the u* completeness level to
$\sim$25.5. Given that the u*-R color in our data is of the order of
1.5, this u* completeness level is fully consistent with the previously
estimated R completeness level (R$\sim$24).

\begin{figure*}
\centering
\mbox{\psfig{figure=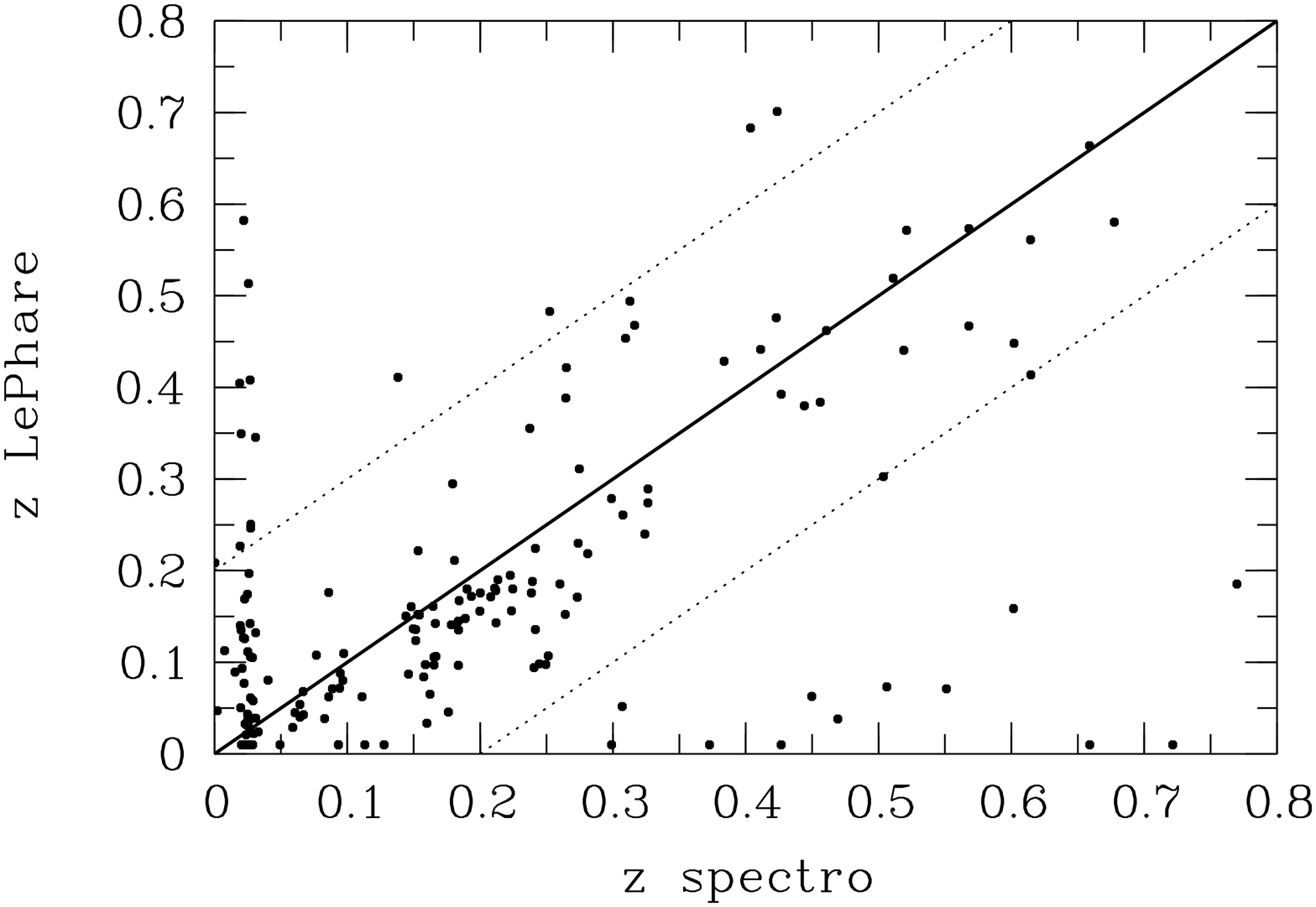,height=6cm,angle=270}\psfig{figure=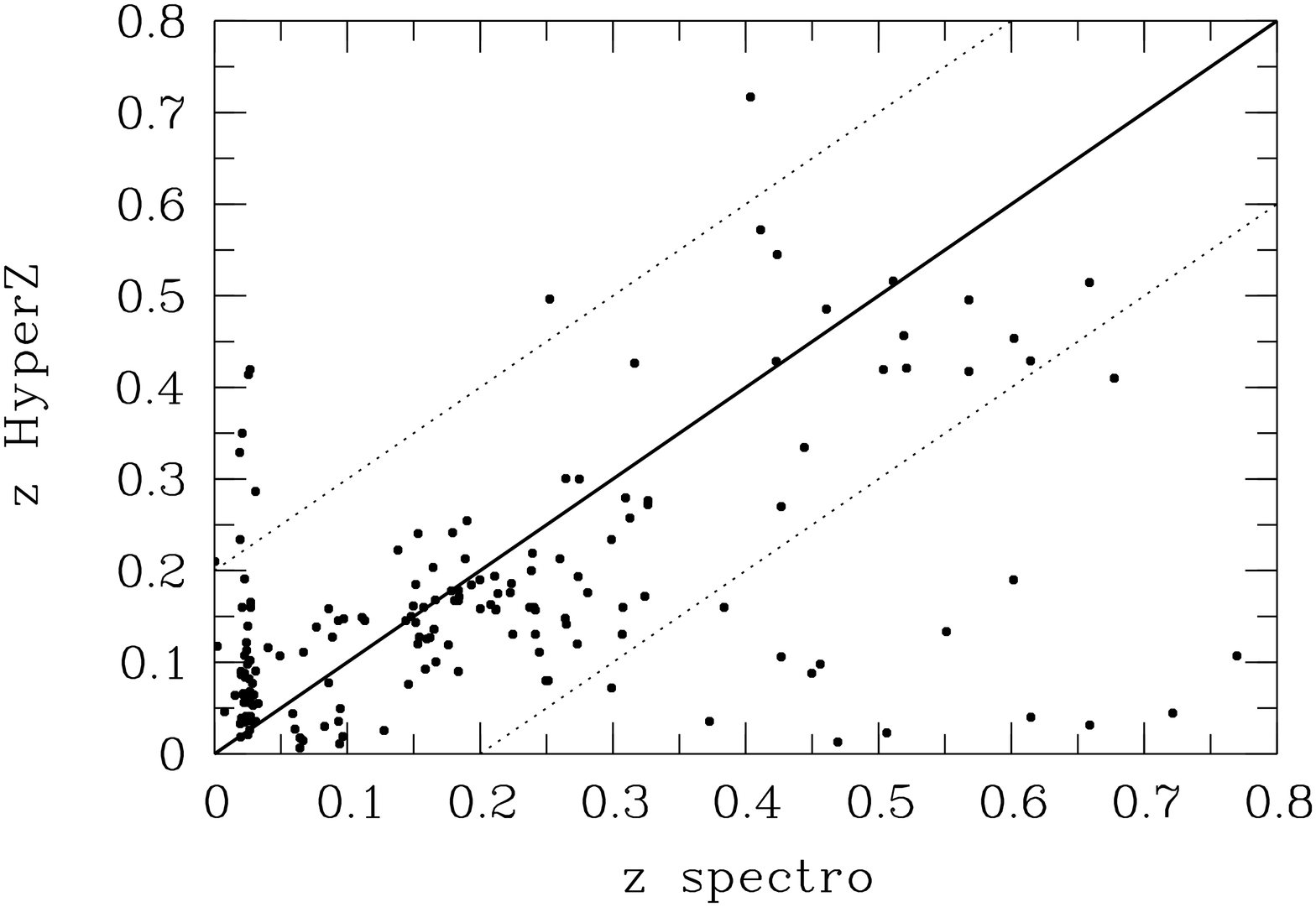,height=6cm,angle=270}}
\caption[]{$LePhare$ (left) and $HyperZ$ (right) photometric redshifts in the CFH12K f.o.v. and
outside the masked regions versus spectroscopic redshifts. The two
dashed lines show a $\pm$0.2 uncertainty enveloppe around the perfect
relation (continuous line). The large number of dots below z$_{phot}$$\sim$0.2
but outside of Coma is explained by the groups detected in section 3.2.}
\label{fig:zz}
\end{figure*}

\subsection{Photometric redshift techniques}

We have computed photometric redshifts with two of the most well-known
packages: $Hyperz$ (e.g. Bolzonella et al. 2000 and
\footnote{http://webast.obs-mip.fr/hyperz/}) and $LePhare$ (e.g. Ilbert
et al. 2006a and
\footnote{http://www.oamp.fr/people/arnouts/LE\_PHARE.html} ({\bf authors:} S. Arnouts
\& O. Ilbert)), mainly to check that these two softwares give consistent
results. These tools are ully described in the quoted
papers. In a few words, Hyperz (Bolzonella et al. 2000) and $Le
Phare$ (Arnouts $\&$ Ilbert) are both based on a template-fitting
procedure. However, the setting of the two photo-z codes presents two
main differences: Hyperz includes a large set of templates from Bruzual
$\&$ Charlot (2003) with different star formation histories and
different ages. $Le Phare$ includes a limited library of 9 templates from
Polletta et al.  (2007) and 3 star-forming templates from Bruzual $\&$
Charlot (2003).  In Hyperz, the zero-point calibration has been done by
comparing the stellar locus in a color-color diagram with the expected
star colors. In $Le Phare$, the zero-point calibration has been done with
an iterative procedure described in section 4.1 of Ilbert et al. (2006a),
by comparing the predicted magnitudes from the best fit template and the
observed magnitudes. In order to compute these possible small magnitude zero
point shifts in $LePhare$,  we
trained the photometric redshift estimates with 
the spectroscopic catalog described in Adami et
al. (2005a). This catalog
was supplemented with 20 archive redshifts from the Keck
LRIS multispectrograph (Secker et al. 1998), increasing the total
number of redshifts available in the CFH12K f.o.v. by $\sim$5$\%$. These
galaxies have magnitudes between R=16.5 and 21.5 and redshifts lower
than 0.5 (only one of these objects is part of the Coma cluster).

The additional zero point shifts applied to our photometry
to compute optimal
photometric redshifts with Hyperz and $LePhare$ were quite small. $Hyperz$ required manually imposed shifts of less
than 0.05 magnitudes in R and I and no shift in u*, B and 
V. $LePhare$ required shifts of 0.028$\pm$0.02 in u*, -0.141$\pm$0.09 in B, -0.062$\pm$0.03 in V,
0.$\pm$0.04 in R and 0.12$\pm$0.05 in I (error bars coming from the dispersion
of the shifts over the whole spectroscopic sample).

A set of fit parameters was then produced (see also Adami et al. 2008): 
internal extinction, age,
multicolor type (based on a color classification in a 5 magnitude
space), redshift, and PDF estimates. However, we mainly consider
here the (photometric) redshift, the template multicolor type, and
the PDF estimates.  Types 1 to 4 are the Elliptical, Sbc, Scd and Irr
from Coleman et al. (1980) respectively and types 5 are starburst 1 templates from
Kinney et al. (1996).

We also computed the integrated probability for each galaxy to
be at a redshift lower than a given value z$_{lim}$ (P hereafter). This
is the area of the PDF enclosed in the considered redshift interval:
[0.,z$_{lim}$].

\subsection{Advantages of the photometric redshift technique}

We could argue that a simple color-magnitude relation
(CMR hereafter) could also provide a good way to discriminate between
cluster members and non members (Biviano et al. 1996). However, besides the
simple fact that five bands appear intuitively better than two to
characterize the spectral energy distributions of galaxies, the use of a
CMR first requires the existence of a well defined red sequence (RS
hereafter). This is not always verified even if the RS in Coma is very
well known at bright magnitudes, thanks to the large percentage of
bright elliptical galaxies in this cluster (e.g. Adami et al. 1998).  In
this case, the photometric redshift and CMR techniques give
comparable results. 

Second, this RS is only poorly known beyond the
present spectroscopic limit (typically R$\sim 21$ for the Coma cluster).
It may show a reddening at R$>21$, as suggested by e.g. Adami et al. (2000).  
Moreover, at
R$>21$ peculiar galaxy types appear, directly resulting from the
disruptions of larger galaxies. An example are the tidal dwarf galaxies (Bournaud et
al. 2003). These peculiar galaxies (not only tidal dwarf galaxies) sometimes have elliptical like
spectral types, but are atypically red (e.g. depleted early type galaxy
cores shown in Adami et al. 2006b). The simple extrapolation of the
bright Coma cluster galaxy RS at R$>21$ would miss these peculiar
objects as cluster members. 

Third, the photometric redshift technique
offers a much more direct way to characterize the spectral energy
distributions of the galaxies. The use of the CMR RS can at best
discriminate between early and late type galaxies for not too faint
magnitudes. On the other hand, the photometric redshift technique (also
see section 2.8)
allows to separate galaxies in 5 spectral types and gives a direct
estimate of the ages and internal extinctions of the galaxies. 

For all these reasons, we decided to apply the photometric redshift
technique in this paper.

\subsection{Results}

First, we limited the catalog to the objects detected in all u*, B, V,
R, and I bands. This excluded the south-west area for which B and V
data are not available (see Adami et al. 2006a).

The two sets of estimates (from $LePhare$ and $Hyperz$) are most of the
time in good agreement.  As usual, we limited the sample to unmasked
regions (this leads to remove all objects at less than twice the
radius of objects brighter than R=18) to avoid uncertain magnitude
estimates, and we plotted for both methods the photometric versus
spectroscopic redshifts. Our spectroscopic catalog includes 172
spectroscopic redshifts in non masked areas, among which 103 are at
z$\leq$0.2. Fig.~\ref{fig:zz} shows the results for $LePhare$ and
$Hyperz$. As both methods provide very similar results, we merged the
two estimates ($LePhare$ and $Hyperz$) keeping the value with the best
reduced $\chi ^2$ (after having homogenized the two sets of $\chi ^2$).

Beyond the general agreement between photometric and spectroscopic
redshifts, we clearly see a degeneracy at the Coma cluster
redshift. For some galaxies which have spectroscopic redshifts
inside the cluster, $LePhare$ and $Hyperz$ produce photometric 
redshifts not only at the
cluster redshift but spread over the interval z=0, z$\sim$0.2. 
Moreover, there is a clear systematic offset of the
order of 0.05 in redshift (mainly visible in the $LePhare$ results)
between z=0.1 and 0.4. This bias can have two origins. First, the
magnitude estimates can be partially biased because of observational
effects (small seeing variations between the different
photometric bands, peculiar problems for a given band at a given sky
location, etc.). Second, cluster galaxies are in general redder than
their equivalents in the field due to environmental effects, while
photometric redshift estimates are computed with synthetic galaxy 
templates mainly based on field galaxies. This can lead to a
mis-interpretation of the spectral energy distribution, confusing the
intrinsic red color and the reddening of galaxies due to redshift. This
shows the need for the definition of synthetic cluster galaxy
spectroscopic templates. This also implies that we cannot use directly the
photometric redshift estimates to efficiently discriminate between
cluster members and field galaxies.

\begin{figure}
\centering \mbox{\psfig{figure=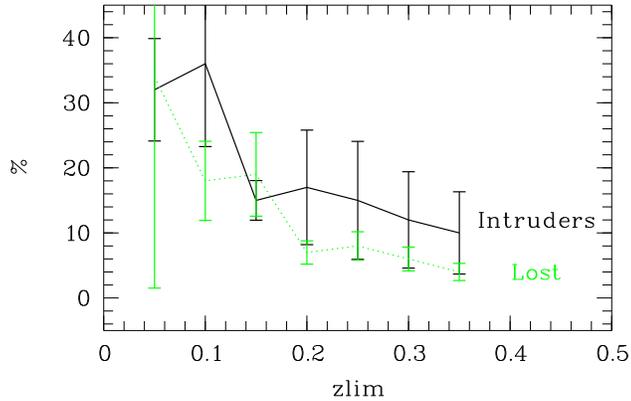,height=6cm,angle=270}}
\caption[]{Variation of the percentage of intruders (full black line)
and lost galaxies (green (grey in black and white version) dotted line)
as a function of z$_{lim}$. Error bars are poissonian uncertainties.}
\label{fig:statbest}
\end{figure}

Moreover, certain galaxies may have a very extended PDF and considering just
the value giving the maximum probability could
produce a wrong photometric redshift. We will therefore classify galaxies as being at
a greater or lower redshift than a limiting value z$_{lim}$, based 
on the probability P quoted above and computed using the area below the
PDF for z lower than z$_{lim}$ for a given galaxy. The value of
z$_{lim}$ will be chosen to minimize the number of intruders (galaxies
at spectroscopic redshift greater than z$_{lim}$ that have a 
photometric redshift lower
than z$_{lim}$) and of lost galaxies (galaxies at spectroscopic redshift lower than
z$_{lim}$ with a photometric redshift greater than z$_{lim}$). This
minimization was done on the basis of the spectroscopic catalog and the
results are shown in Fig.~\ref{fig:statbest}.  Setting z$_{lim}$ to a
high value would produce no intruder or lost galaxies, but the
discrimination power of such a high value would be low. We therefore
selected z$_{lim}$=0.20, which is a good compromise between intruder and
lost galaxy percentages and the discrimination power of the method. We
see in Fig.~\ref{fig:statbest} that the percentages of intruder or lost
galaxies do not decrease very strongly for z$\geq$0.2.  With this limit
and considering our spectroscopic catalog, we estimate that we lose (in
the magnitude range covered by the spectroscopic catalog) less than
10$\%$ galaxies and we include about 15$\%$ intruders. These percentages
will be taken into account when computing the GLFs in the next section.

\subsection{Uncertainties in the galaxy discrimination based on the PDF}

With this limit of z=0.2 and considering our spectroscopic catalog, we
estimate that we lose less than 10$\%$ galaxies and we include about
15$\%$ intruder galaxies if we limit our analysis to the spectroscopic
catalog magnitude limit.

\begin{figure}
\centering \mbox{\psfig{figure=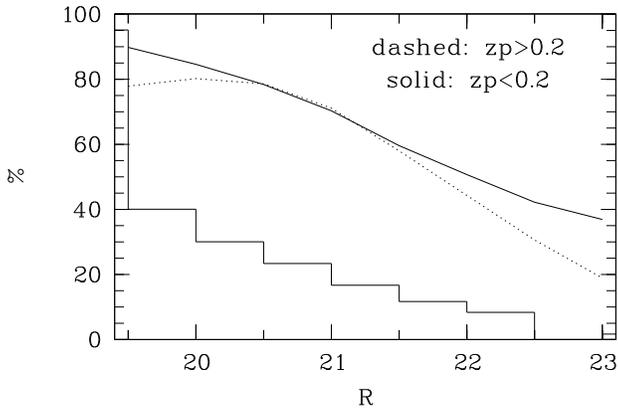,height=6cm,angle=270}}
\caption[]{Fraction of photometric redshifts (as a function of R
magnitude) from $LePhare$ with a $1\sigma$ error bar smaller than
$0.2\times (1+z)$. Solid line: photometric redshifts lower than 0.2,
dashed line: photometric redshifts greater than 0.2. The histogram (in
arbitrary units) shows the number of available spectroscopic redshifts
as a function of R magnitude.}  \label{fig:faintspectro}
\end{figure}

Given the fact that our spectroscopic catalog is by far not as deep as
our photometric catalog, we need to quantify the accuracy of the
photometric redshifts beyond the spectroscopic limit. Ilbert et
al. (2006a) have shown that the $1\sigma$ error bars are representative
of a measurement at 68$\%$ confidence level, and we can therefore
quantify this accuracy based on the 1$\sigma$ error
bars. Fig.~\ref{fig:faintspectro} shows the fraction of photometric
redshifts from $LePhare$ (results would be similar with Hyperz) with a
$1\sigma$ error bar smaller than $0.2\times (1+z)$ (0.2 being typically
the width of the low redshift interval we want to characterize). This
figure shows that we still have more than half of the galaxies with a
photometric redshift estimate lower than $0.2\times (1+z)$ for R
brighter than 22. We also have a good agreement between the decrease of
the fraction of photometric redshifts with a $1\sigma$ error bar smaller
than $0.2\times (1+z)$ and the number of available spectroscopic
redshifts (shown as the histogram in
Fig.~\ref{fig:faintspectro}). These percentages will be taken into
account when computing the GLF uncertainties in the next section.

\subsection{Uncertainties in the galaxy color-type estimate}

      In order to determine the reliability of color-types assigned to
galaxies in the Coma cluster, we have carried out a series of simulations
with {\it Hyperz} and related software. Synthetic catalogs contain
10$^5$ galaxies in total at the redshift of Coma, with R-band magnitudes
ranging between R$=$19 and 24, spanning all the basic spectral types, 
i.e. $\sim$4000 galaxies per magnitude and type bin. 
Photometric errors in the different filters were assigned following a
gaussian distribution with variance scaled to apparent magnitude, i.e.
$\sigma(m) \simeq 2.5 \log [1+1/(S/N)]$, where $S/N$ is the signal to
noise ratio corresponding to the apparent magnitude $m$ in our catalog.
{\it Hyperz} settings used to fit these synthetic catalogs are the
same as for real catalogs. 

    Table~\ref{types_simulated} summarizes the results obtained from
these simulations as a function of R-band magnitude and photometric
type. Up to R$\le$23 (we assumed a limit of R=22.5 in the following),
photometric types are correctly retrieved by SED fitting, with a
percentage of failures usually below $\sim 3\%$, and up to $\sim 10\%$
for the bluest galaxies. As expected, early-type galaxies are better
identified than late types, even for the faintest sources in our
catalog, but the difference is small. At least $\sim 75\%$ of galaxies
are still correctly classified in the faintest magnitude bin.

\begin{table}
\caption{Percentage of simulated galaxies in the Coma cluster with
photometric types correctly assigned as a function of R-band magnitude. 
}
\begin{center}
\begin{tabular}{cccccc}
\hline
R   &  E/S0 & Sbc & Scd & Im & SB \\
mag &  \%   & \%  & \%  & \% & \% \\
\hline
19-20  &   100.0 &   99.9  &   99.9  &   98.9  &   98.8 \\
20-21  &   100.0 &   99.9  &   99.9  &   98.7  &   99.2 \\
21-22  &   100.0  &  99.7  &   99.8  &   97.9  &   97.5 \\
22-23  &   100.0 &   98.3  &   98.2  &   93.3  &   90.3 \\
23-24  &    98.0 &   88.4  &   84.7  &   76.2  &   77.3 \\
\hline
\end{tabular}
\label{types_simulated}
\end{center}
\end{table}

Of course, since we fit the templates used to generate the
  catalogues, this must help recovering the color-types; the
  {\bf  errors quoted are therefore lower limits on the true uncertainties}
  on the galaxy color-type estimates (see also Section 5).

\section{The Galaxy Luminosity Function computation}

As previously stated, photometric redshift techniques allow an
optimal subtraction of background galaxies for a redshift limit of 0.2.
However, we still need to remove from the sample the galaxies at
z$\leq$0.2 which do not belong to the Coma cluster. These galaxies can
be field galaxies or galaxies included in groups of galaxies not related
to Coma. We chose to subtract statistically these two contributions by
considering field and group luminosity functions taken in the
literature. In this way, we are still applying a statistical
subtraction, but photometric redshift techniques help to cut
down very significantly the redshift range that includes background
galaxies.

\subsection{Field contribution}

We estimated the field contribution to be subtracted from the VVDS field luminosity
function of Ilbert et al. (2005), who obtained a very large
redshift catalog complete to I(AB)$\sim 24$ (close to R$\sim 24$
considering the passband of our filters). 
This is close to our own magnitude limit and ensures that the
field luminosity function we subtract is really constrained over our
whole magnitude range, without requiring any extrapolation.

The Ilbert et al. (2005) luminosity function is also computed in five
bands from U to I, allowing a field subtraction adapted to each
photometric band.

The subtraction of the field contribution was simply done by computing
the cosmological co-moving volume included in our field of view at redshift
lower than 0.2 and using the $\phi ^*$, $\alpha$, and M$^*$ parameters
of the Schechter function given by Ilbert et al. (2005).

\subsection{Group contribution}

We first located in our field of view the groups unrelated with the
Coma cluster and at redshift lower than 0.2 by applying the
Serna-Gerbal method (Serna \& Gerbal 1996) to the redshift catalog
(Adami et al. 2005a). Briefly, this method is able to
detect dynamically linked galaxies (what we call a group of galaxies)
based on the positions, magnitudes and redshifts of the galaxies (see
Table~\ref{tab:group}). This method was already applied by Adami et
al. (2005a) to study the Coma cluster, and five background groups at
z$\leq$0.2 were detected.

\begin{table}
\caption{Characteristics of the background groups detected by the
  Serna-Gerbal method, based on spectroscopic redshifts: coordinates,
  mean redshift and total mass (Adami et al. 2005).}
\begin{center}
\begin{tabular}{lllll}
\hline
 Group  & $\alpha$ & $\delta$   &  mean z & mass ( M$_{\odot}$) \\ 
\hline
G1 & 195.04 & 27.61 & 0.149 & 7.34 10$^{12}$ \\
G2 & 194.90 & 27.71 & 0.138 & 5.25 10$^{11}$ \\
G3 & 194.92 & 27.87 & 0.144 & 1.25 10$^{13}$ \\
G4 & 194.94 & 28.27 & 0.097 & 1.46 10$^{11}$ \\
G5 & 194.70 & 27.89 & 0.133 & 4.15 10$^{12}$ \\
\hline
\end{tabular}
\end{center}
\label{tab:group}
\end{table}

\begin{figure*}
\centering
\mbox{\psfig{figure=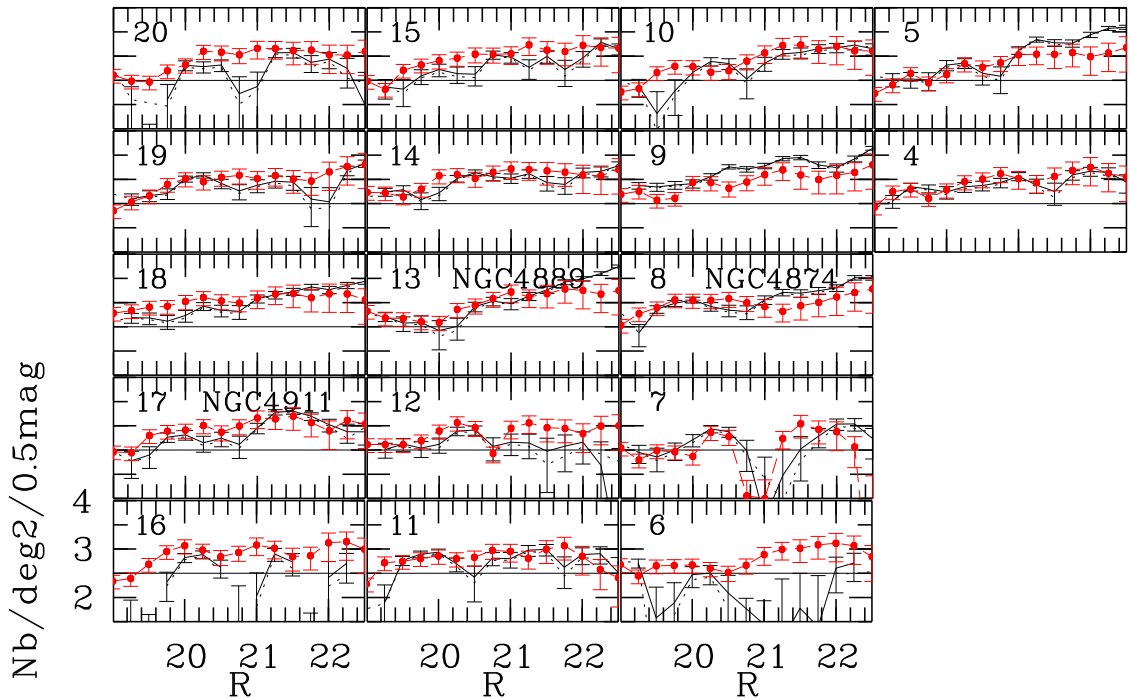,height=12cm,angle=270}}
\caption[]{R band GLFs for different regions in the Coma cluster. North
is top and east is left. The empty subgraphs correspond to areas where B
and V data were not available. Black continuous and short dashed lines
(along with error bars) are the GLF statistical estimates of Adami et
al. (2007a). Red (grey in black and white version) long-dashed lines
(along with error bars) are the present estimates. }
\label{fig:fdlzphot}
\end{figure*}

The group luminosity functions were computed from the SDSS group
luminosity function estimates of Zandivarez et al. (2006). Since these
authors have data in the five SDSS bands, we transformed their
magnitudes to our own system applying the relations given by Fukugita et
al. (1995). The number of galaxies brighter than R=17.75 (where our
spectroscopic catalog is nearly complete) was estimated from our
spectroscopic catalog. This allowed us to compute the $\phi ^*$
normalisation of the five detected groups (see Fig.~\ref{fig:imageu}).

In order to subtract this group contribution, we assumed that the
whole group galaxy contribution was enclosed in a 300 kpc diameter
circle, a typical group size. 

We note here that we do not know the background group GLF per spectral
type and this prevents us from computing a Coma cluster GLF
$per~spectral~type$ with the photometric redshift technique. We will
simply study the general multicolor type spatial distribution in
Section~5.

\subsection{GLF uncertainties}

We took into account several uncertainties on the Coma cluster GLF
(computed in 0.5 magnitude bins, as in Adami et al. 2007a):

- the Poisson noise in each bin;

- the uncertainty on the field and group luminosity functions,
computed by generating 1000 field and group luminosity functions with
parameters gaussianly varying inside the error range quoted
in the literature;

- the uncertainty due to the photometric redshift computations. As
previously shown, the number of galaxies below z=0.2 can be
overestimated by 15\% or underestimated by 10$\%$ within the magnitude
range where the spectroscopic catalog is contributing. For
fainter magnitudes, we considered Fig.~\ref{fig:faintspectro} to infer a
value of the overestimation and of the underestimation. 

The sum of the three uncertainties is assumed to be the total error and 
is close to a 3$\sigma$ error. The uncertainty due to the photometric redshift
computations is clearly the dominating source of error.

\section{Galaxy Luminosity Functions (GLF)}

\subsection{Galaxy Luminosity Functions based on photometric redshifts}

We computed the GLF in the subregions previously considered by Adami et
al. (2007a).  These $\sim$10'$\times$10' regions cover the whole cluster
area and represent a good compromise between the spatial resolution and
the number of galaxies included in the individual GLFs.
Fig.~\ref{fig:fdlzphot} shows the GLFs together with their 1$\sigma$
error bars (red symbols) overplotted on the previous GLFs computed by
Adami et al. (2007a: black symbols) with statistical background
subtractions.

The goal of this section is mainly to compare our results with those of
Adami et al. (2007a). As these two studies are based on the same
dataset, this is a good way to test the reliability of the two methods
(statistical field subtraction and photometric redshift technique). The
agreement is generally quite good between the two estimates. A few
regions, however, show significant discrepancies: fields
5, 6, 8, 9, 16, and 20. These fields have 
different luminosity functions (derived with the two
techniques) over more than 15$\%$ of the magnitude interval R=[19.25, 22.5],
the largest differences occuring for fields 16 and 6. 
It is interesting to note that among these fields,
three contain very bright stars (fields 5, 8, and 9) and one (field 8) the
cluster dominant galaxy NGC~4874. These objects have a very extended light halo
that possibly affects the photometry (even if they were masked inside twice 
their radius) and therefore the photometric
redshift computations, or the faint object detection in the case of the statistical
background removal technique of Adami et al. 2007a). 
We also know that at least one of these fields (field 9)
contains a significant population of very faint and very blue Coma galaxies (the
field around NGC~4858/4860, Adami et al. 2007b). These faint blue
knots, similar to those discussed by Cortese et al. (2007), are
perhaps not well represented in our galaxy synthetic templates; this
could produce an incorrect value for photometric redshifts, outside the
Coma cluster range. 

\begin{figure}
\centering
\mbox{\psfig{figure=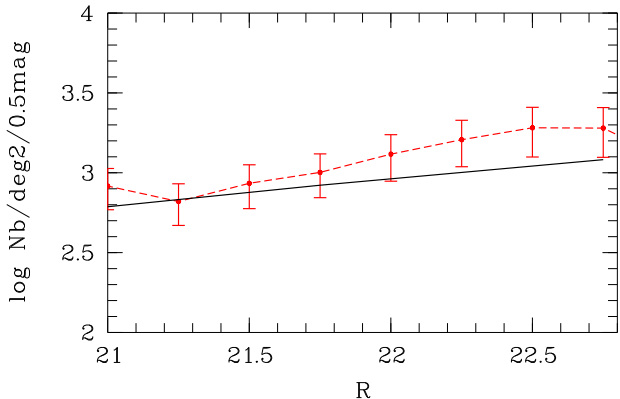,height=6cm,angle=270}}
\caption[]{R band GLF for the NGC~4874 field computed with photometric
redshifts (red dashed line - grey in black and white version). The slope
from Secker et al. (1996) is overplotted (black full line). }
\label{fig:secker}
\end{figure}

In order to have a third estimate (beyond Adami et al. 2007a and 
the present paper determinations), we also compared our results for one
of these possibly biased fields (the NGC~4874 region) with a study
dedicated to the computation of the GLF in the Coma cluster center
(Secker et al. 1996, based on Keck images). Fig.~\ref{fig:secker} shows that
these authors predict galaxy number densities consistent at the
2$\sigma$ level with the photometric redshift technique estimates in the
range where these counts are lower than the ones computed in Adami et
al. (2007a). So the counts of Adami et al. (2007a) could also be
overestimated (despite the great care that was devoted to the
statistical field subtraction). It is beyond the scope of this paper to
present a full comparison with literature results. This was already done
in Adami et al. (2007a: section 7.4).

\begin{figure}
\centering
\mbox{\psfig{figure=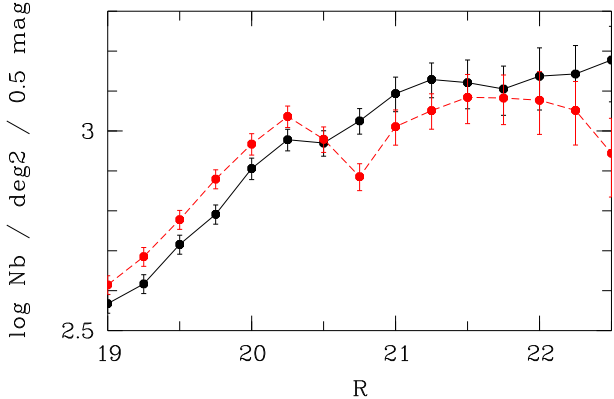,height=6cm,angle=270}}
\caption[]{R band GLF for the northern (black) and southern (red dashed line - 
grey in black and white version) parts as defined in Adami et
al. (2007a). } \label{fig:nordsud}
\end{figure}

We also note that we still see the discrimination between the
north-northeast and the south-southwest parts of the cluster
(Fig.~\ref{fig:nordsud}) that was detected by Adami et al. (2007a). The
north-northeast region is more populated in the faint magnitude regime
than the south-southwest regions. However, this trend is not significant
when applying the photometric redshift technique. We also see in the
southern cluster part a similar turn-over as the one already shown in R
in Adami et al. (2007a), but it is poorly significant due to
the large error bars. These error bars could only be reduced by using a
deeper spectroscopic catalog to better constrain the photometric redshift estimates
of the faint galaxies.

\subsection{Galaxy Luminosity Functions in the  u* band 
based on statistical background subtraction}

The Megacam u* band image is more extended than the CFH12K f.o.v. We
typically have a 7~arcmin strip all around the CFH12K f.o.v. which is
only covered in u*. In order to compute u* GLFs in this external area,
we applied the same statistical field galaxy subtraction technique as in
Adami et al. (2007a and b). The main goal of this computation was to
investigate potential variations of the GLF faint end slope in the u*
band, which is sensitive to recent star formation bursts, and to compare our
results to those of Donas et al. (1995).

The comparison fields were those among the deep CFHTLS fields (D1, D3, and D4)
that are deep enough to allow a statistical
subtraction from our Coma data. In order to
limit these catalogs to the same depth, we selected only galaxies
brighter than u*=24 in total magnitude and brighter than 26.3 in
surface brightness.

\begin{figure*}
\centering \mbox{\psfig{figure=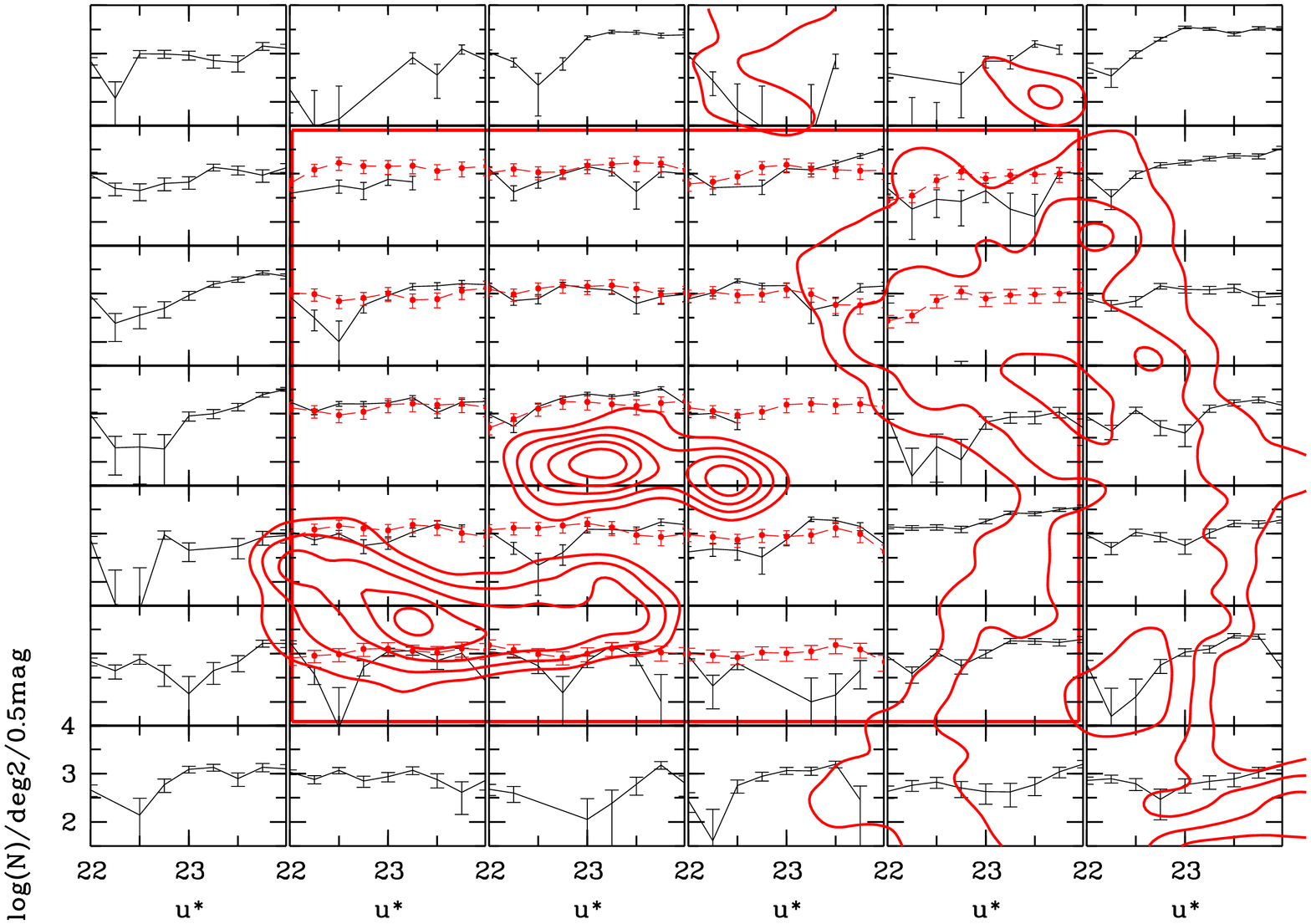,height=10.3cm,angle=270}}
\mbox{\psfig{figure=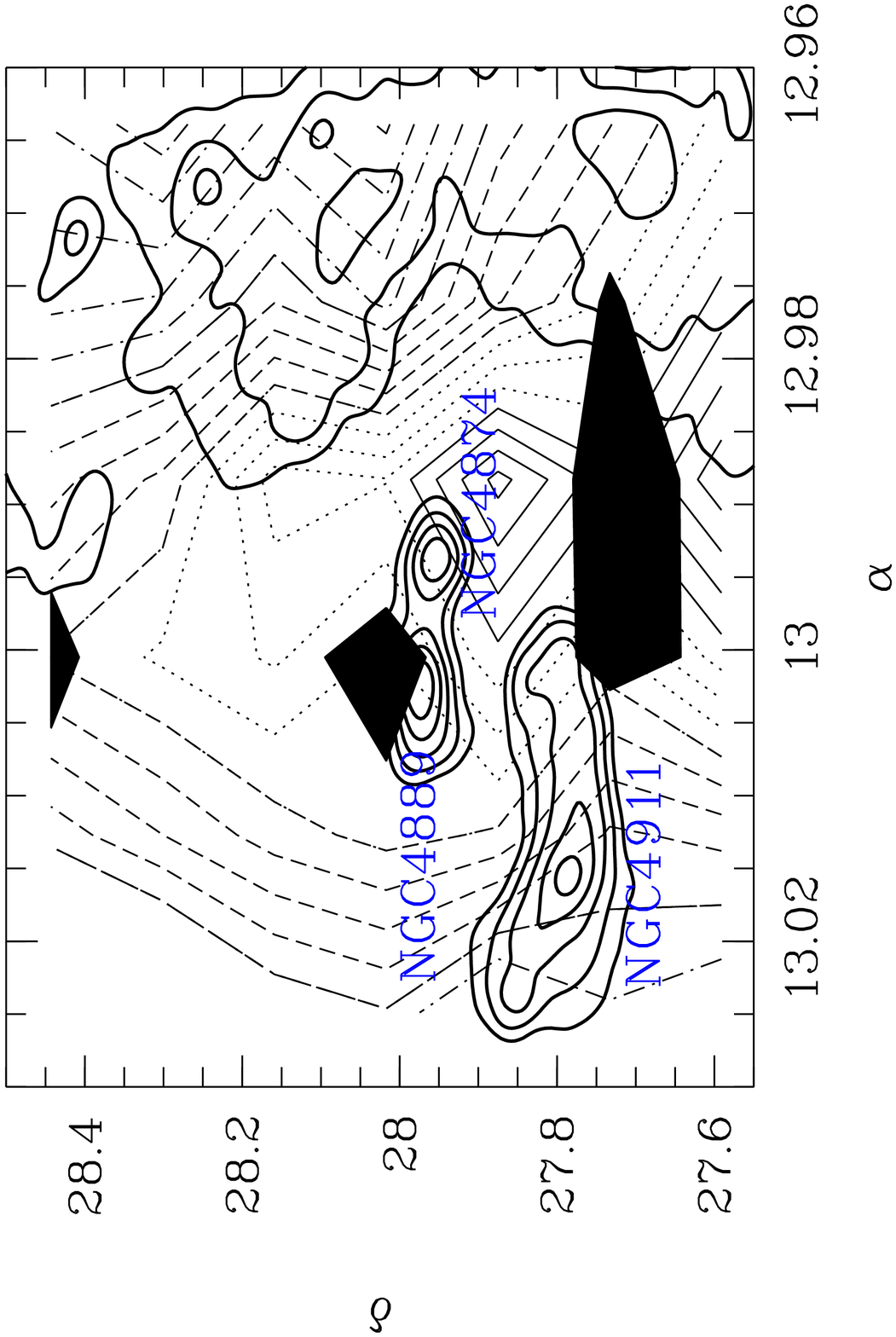,height=10.8cm,angle=270}}
\caption[]{Upper panel: u* band GLFs for different regions in the Coma
cluster computed with a statistical subtraction (continuous black lines
with error bars) and the photometric redshift technique (red (grey
in black and white version) dots linked with dashed lines and error
bars). North is top and east is left. The CFH12K field
of view is shown as the central red (grey in black and white version)
rectangle. The empty subgraphs correspond to areas where the field
counts were greater than the cluster counts. X-ray substructures are
shown as thick continuous contours.  Lower
panel: contours of the GLF slope within the Megacam f.o.v. (thin continuous lines: slopes
between $-1$ and $-1.2$, thin dotted lines: between $-1.2$ and $-1.4$, thin long dashed lines: between
$-1.4$ and $-1.6$, thin long dot-dashed lines: between $-1.6$ and $-2.2$).  Shaded areas
correspond to zones where the slope is not significantly different from
0 at more than a 2$\sigma$ level. X-ray substructures from Neumann et
al. (2003) are shown as smooth thick black contours.}  \label{fig:iso}
\end{figure*}

The comparison fields cover a total area of more than 3~deg$^2$, thus
reducing the cosmic variance (see Adami et al. 2007a).  The resulting
GLFs are similar to the u* band GLFs computed in the CFH12K f.o.v. 
based on
photometric redshifts. We sub-divided the Megacam f.o.v. in 7$\times$6
subfields and computed the slope $\alpha$ of the u* GLF (computed applying
a statistical subtraction and modelled by a Schechter function) in the
u*=[22,24] magnitude range for each zone. The 42 $\alpha$ slopes 
are given in Table~\ref{tab:slopesu} and were
used to generate Fig.~\ref{fig:iso}, where the contours
of GLF slope $\alpha$ within the Megacam f.o.v. are displayed.

\begin{table}
\caption{Slope (and associated 1-$\sigma$ uncertainty) of the u* Schechter luminosity
function between u*=22 and 24 as a function of coordinates. These values were used to generate 
Fig.~\ref{fig:iso}.}
\begin{center}
\begin{tabular}{llll}
\hline
$\alpha$   &  $\delta$ & Slope & Err. slope \\
\hline
12.9639 & 27.5911 & -1.36 & 0.22 \\
12.9639 & 27.7339 & -2.22 & 0.61 \\
12.9639 & 27.8767 & -1.52 & 0.18 \\
12.9639 & 28.0194 & -1.59 & 0.28 \\
12.9639 & 28.1622 & -1.18 & 0.17 \\
12.9639 & 28.3050 & -1.96 & 0.21 \\
12.9639 & 28.4478 & -2.20 & 0.27 \\
12.9761 & 27.5911 & -1.48 & 0.20 \\
12.9761 & 27.7339 & -1.71 & 0.21 \\
12.9761 & 27.8767 & -1.61 & 0.09 \\
12.9761 & 28.0194 & -1.84 & 0.56 \\
12.9761 & 28.1622 & -1.84 & 0.56 \\
12.9761 & 28.3050 & -1.43 & 0.41 \\
12.9761 & 28.4478 & -1.96 & 0.40 \\
12.9883 & 27.5911 & -2.07 & 0.74 \\
12.9883 & 27.7339 & -0.34 & 0.57 \\
12.9883 & 27.8767 & -1.67 & 0.31 \\
12.9883 & 28.0194 & -0.13 & 0.65 \\
12.9883 & 28.1622 & -0.98 & 0.28 \\
12.9883 & 28.3050 & -1.81 & 0.23 \\
12.9883 & 28.4478 & -0.08 & 1.51 \\
13.0006 & 27.5911 & -1.41 & 0.49 \\
13.0006 & 27.7339 & -0.32 & 0.63 \\
13.0006 & 27.8767 & -1.66 & 0.33 \\
13.0006 & 28.0194 & -1.62 & 0.24 \\
13.0006 & 28.1622 & -0.85 & 0.18 \\
13.0006 & 28.3050 & -1.08 & 0.26 \\
13.0006 & 28.4478 & -1.99 & 0.37 \\
13.0128 & 27.5911 & -0.73 & 0.16 \\
13.0128 & 27.7339 & -1.18 & 0.73 \\
13.0128 & 27.8767 & -1.25 & 0.22 \\
13.0128 & 28.0194 & -1.06 & 0.13 \\
13.0128 & 28.1622 & -1.96 & 0.41 \\
13.0128 & 28.3050 & -1.51 & 0.18 \\
13.0128 & 28.4478 & -2.62 & 0.53 \\
13.0250 & 27.5911 & -1.88 & 0.37 \\
13.0250 & 27.7339 & -1.52 & 0.39 \\
13.0250 & 27.8767 & -2.12 & 0.78 \\
13.0250 & 28.0194 & -2.30 & 0.43 \\
13.0250 & 28.1622 & -2.12 & 0.28 \\
13.0250 & 28.3050 & -1.43 & 0.18 \\
13.0250 & 28.4478 & -1.63 & 0.36 \\
\hline
\end{tabular}
\end{center}
\label{tab:slopesu}
\end{table}

We clearly see a steepening of the u* GLF slope between the center and
outskirts, changing from $-1$ to $-2.2$. 
The regions with the steepest u* GLFs correspond to those
where X-ray subtructures are detected and are located at $\sim$25' from the
cluster center. This is exactly where Donas et al. (1995) detected an
enhancement in the median UV flux and in the fraction of bright blue star forming
galaxies considered as cluster members.

\section{Multicolor type spatial distribution}

\begin{figure*}
\centering
\mbox{\psfig{figure=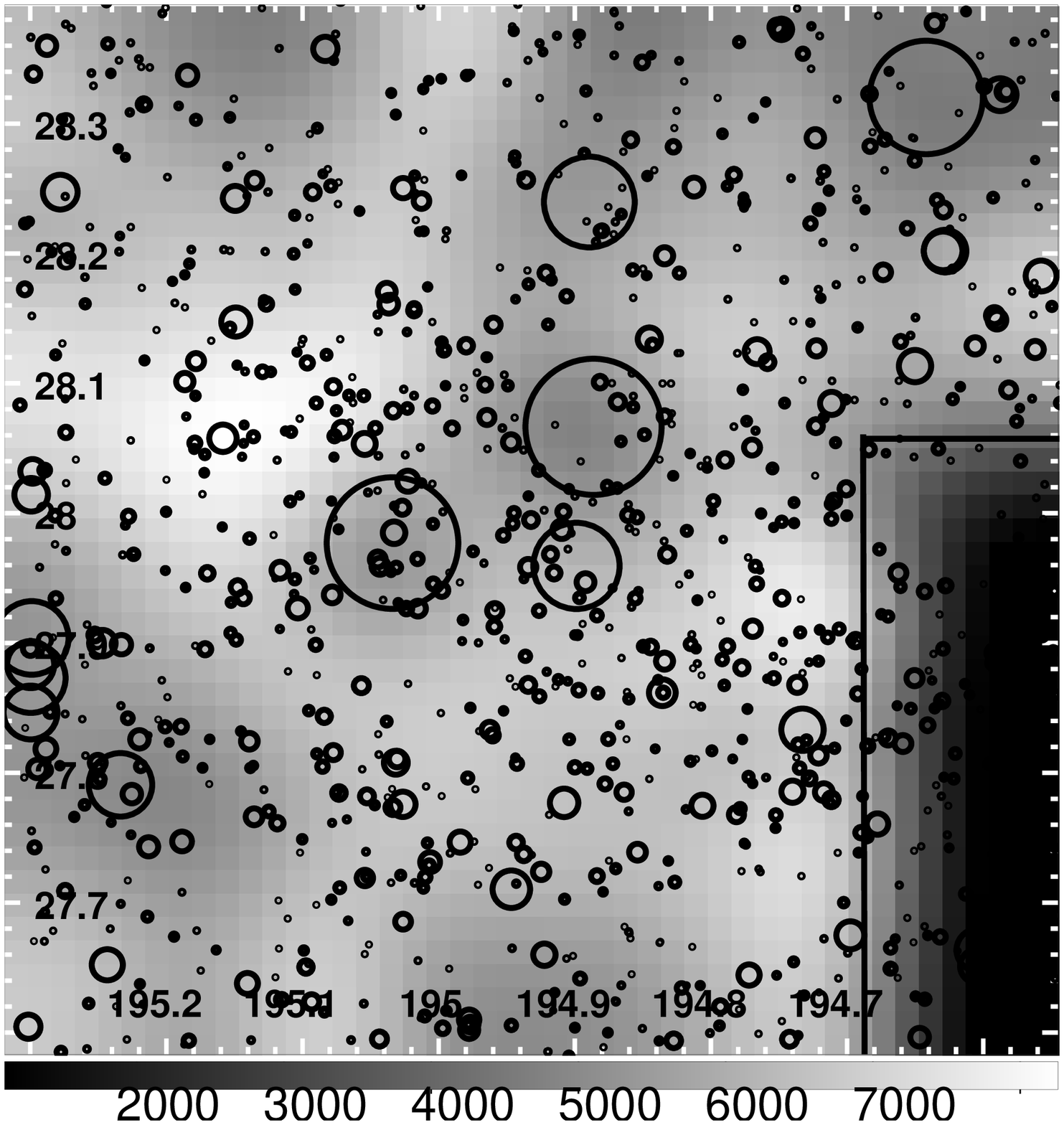,height=6.5cm,angle=0}\psfig{figure=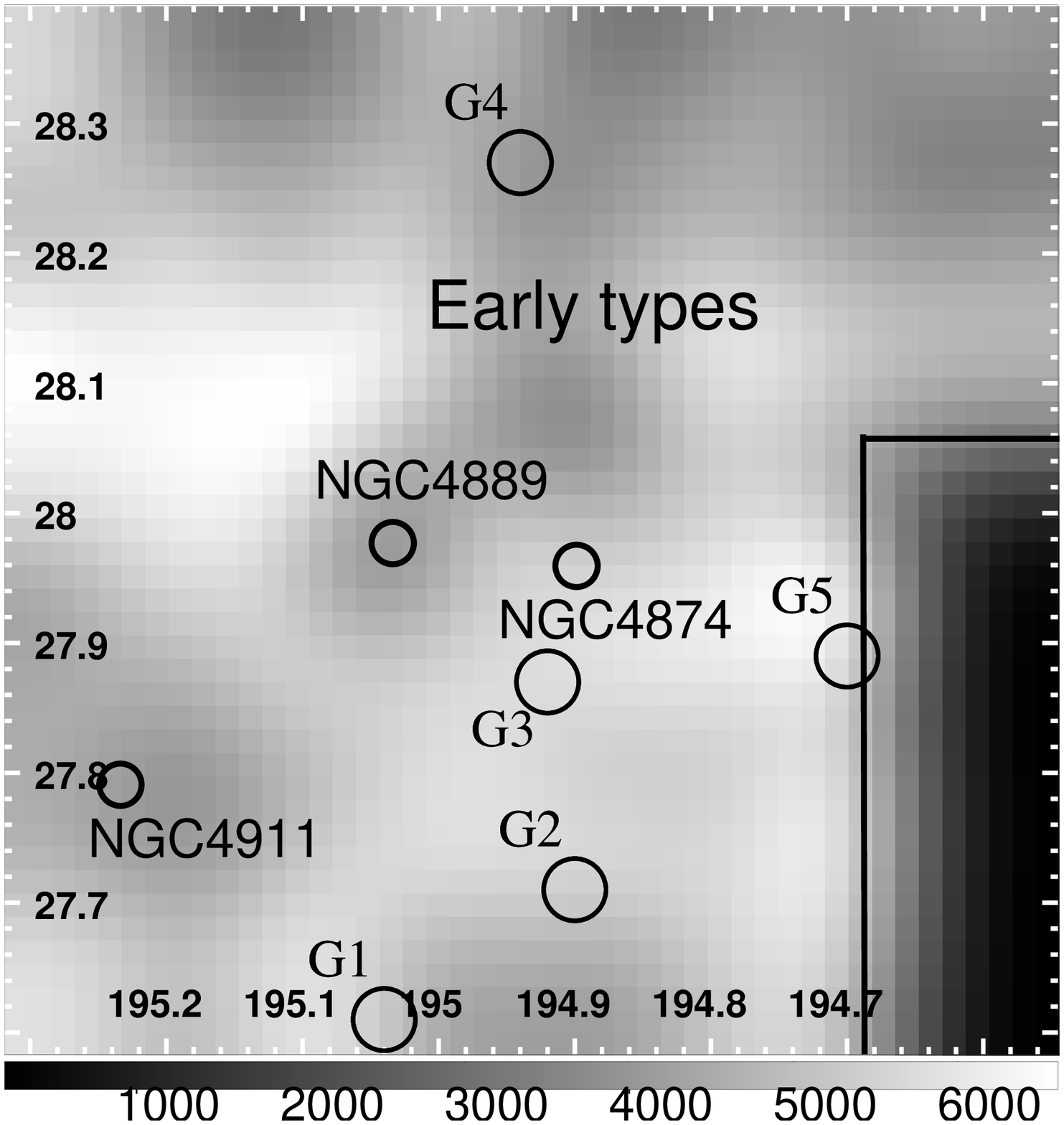,height=6.5cm,angle=0}\psfig{figure=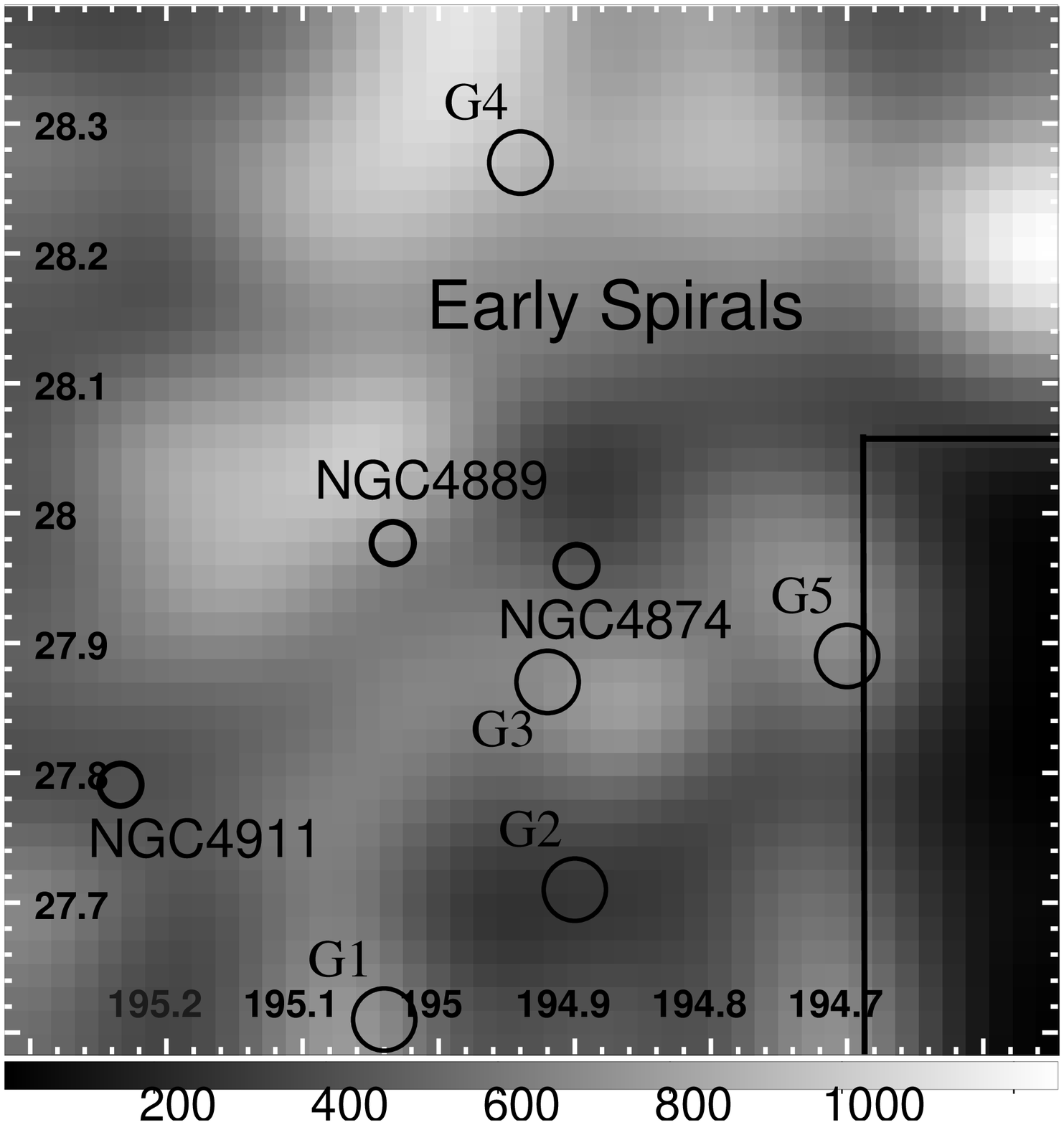,height=6.5cm,angle=0}}
\mbox{\psfig{figure=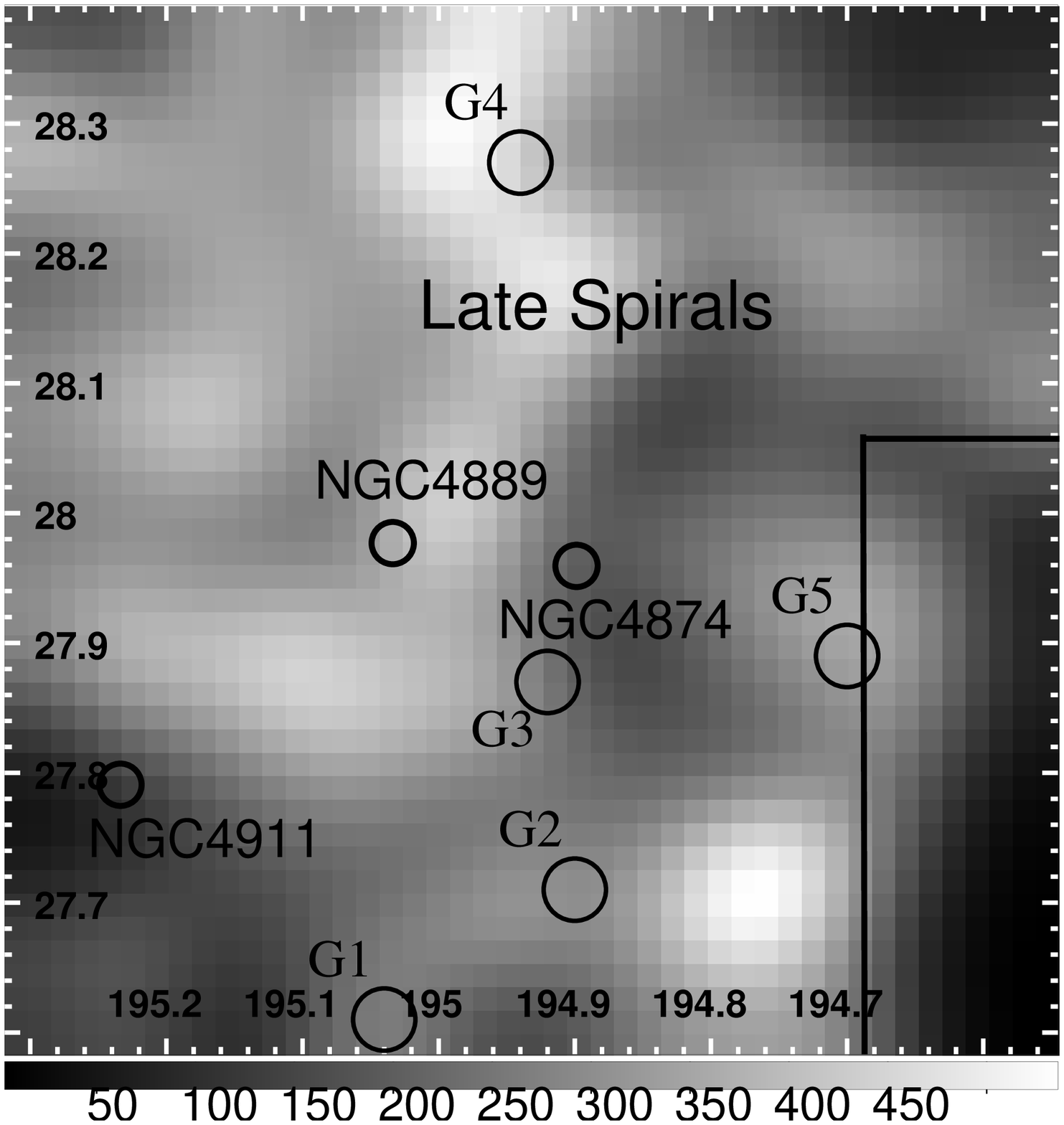,height=6.5cm,angle=0}\psfig{figure=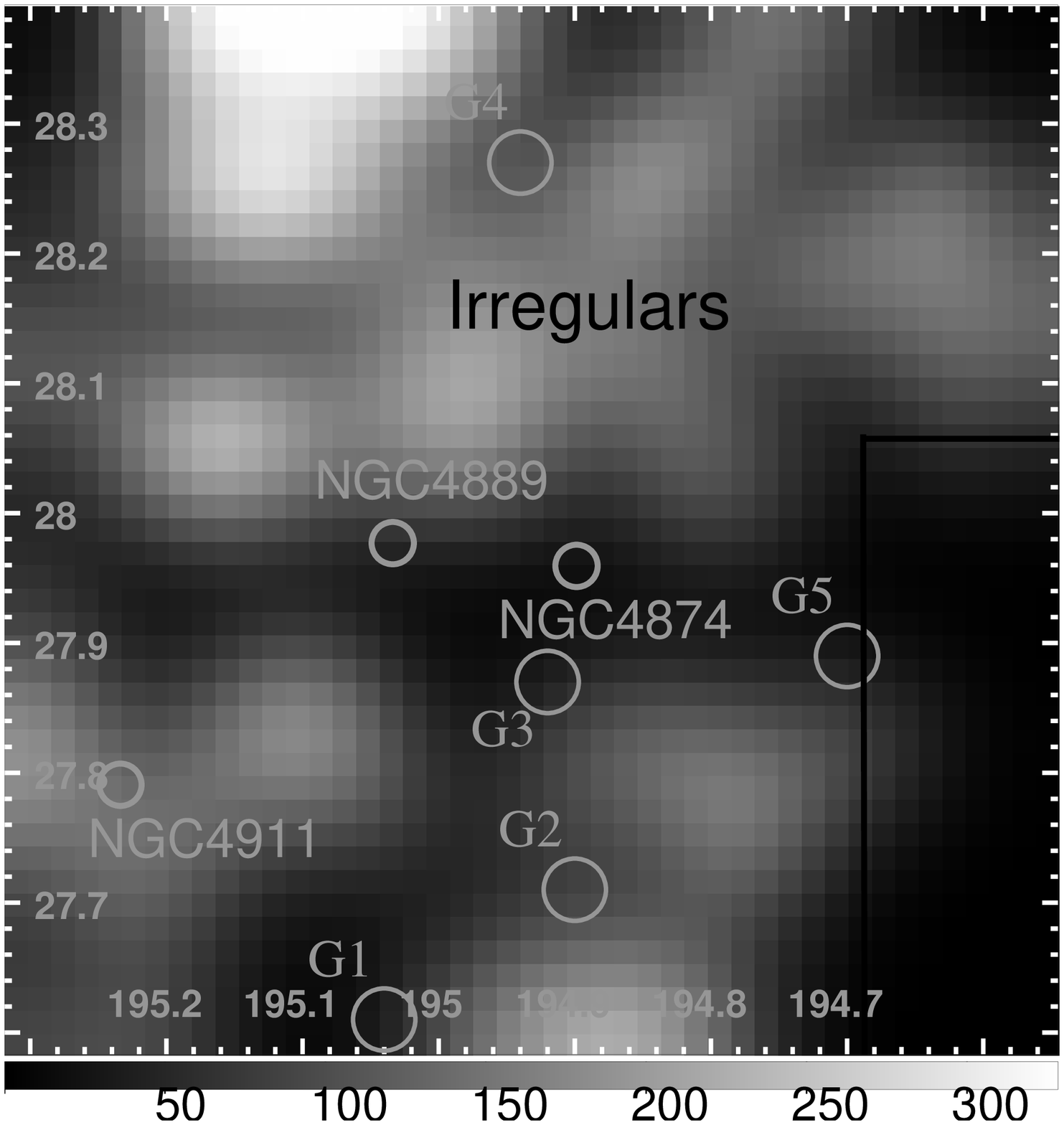,height=6.5cm,angle=0}\psfig{figure=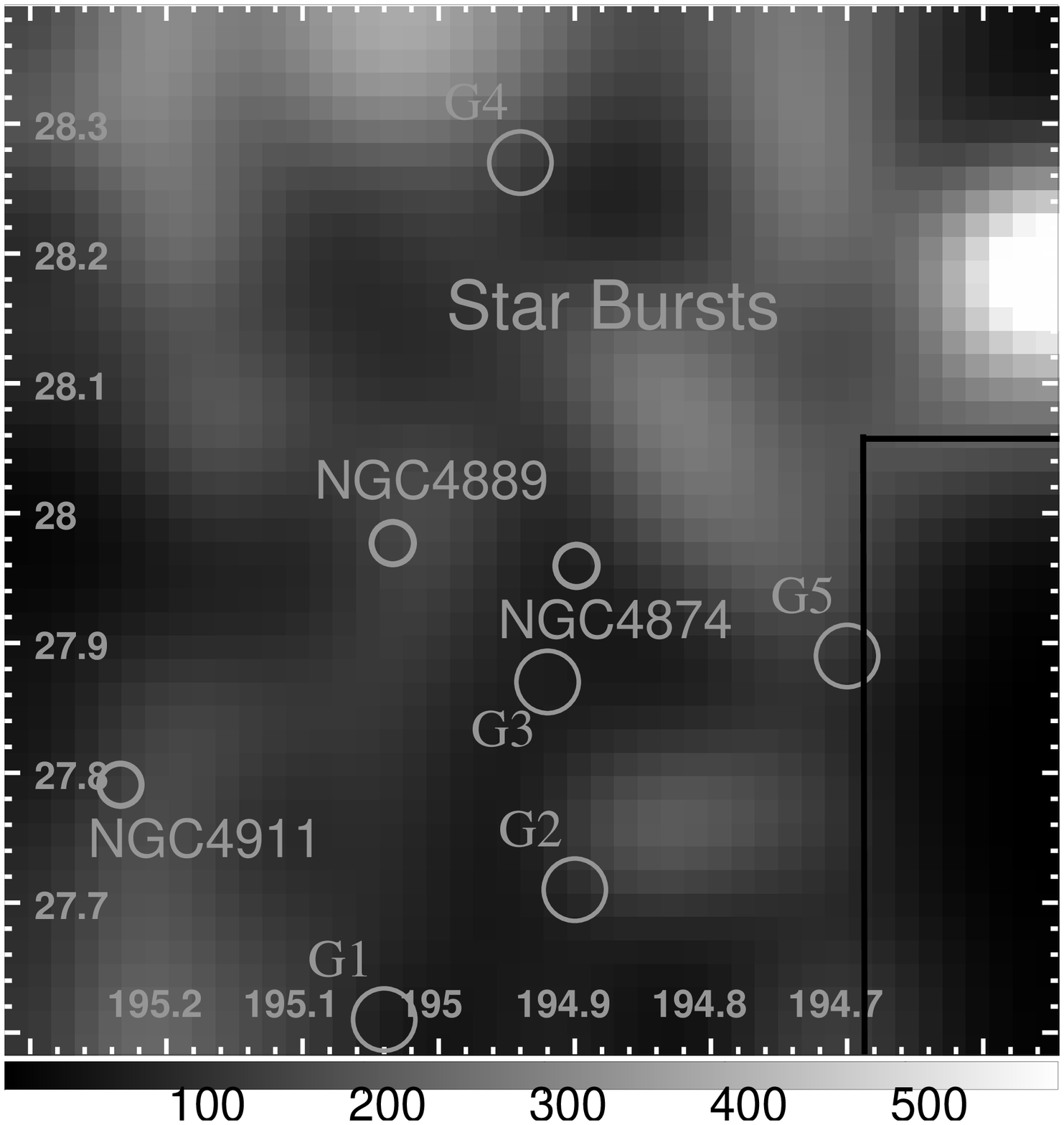,height=6.5cm,angle=0}}
\caption[]{Density contours of the various multicolor galaxy types: upper left: all
multicolor types overplotted with masked regions (circles), upper
middle: early type galaxies (types 1), upper right: early spirals
(types 2), lower left: late spirals (types 3), lower middle:
irregulars (type 4), lower right: starburst galaxies (type 5). The
lighter color in the maps, the highest the number density peak. In each
figure, north is up and east is left, the total field of view is
that of the CFH12K (42$\times$52~arcmin$^2$), and the greyscale levels are
given in number of galaxies per deg$^2$. 
The 5 detected loose groups are indicated as large circles of 300kpc diameter (at
z=0.1). Small circles give the positions of the three brightest galaxies in
the field. {\bf In each of the panels, we finally show the south-west area
  where B and V filters are missing.}}
\label{fig:denspour}
\end{figure*}

The goal of this section is to investigate the spatial distribution of
galaxies according to their multicolor type. We stress here once
again that our multicolor types are based on a color classification in a
5 magnitude space and are not real morphological types. Real
morphological types are already not well determined at R$\geq$18 (our
starting magnitude) and completely unknown at R$\geq$22, waiting for the
results of the Coma cluster HST imaging survey (e.g. Carter et
al. 2008). If we consider the morphological types compiled in Biviano
et al. (1996) for 8 galaxies fainter than R=18 for which we 
computed photometric redshifts, we find that all are classified as 
elliptical
galaxies and all have a multicolor type assigned to Delta bursts or
elliptical galaxies, from the Bruzual $\&$ Charlot (2003) evolutionary
synthetic SEDs. This agreement clearly needs, however, to be confirmed.

Due to masked regions, 
we postpone the galaxy density profile per
multicolor type determination to a future work and limit our analysis
to the search of galaxy concentrations outside the masked areas.

In order to estimate the multicolor type spatial distribution
inside the Coma cluster, we must adress the problem of z$\leq$0.2 non Coma
member galaxies (field or loose group objects) that cannot be discriminated 
by the photometric redshift technique alone. 

If we consider the field galaxy luminosity function computed by Ilbert
et al. (2006b) from similar data, we find that the field contribution
represents about 15$\%$ of the Coma cluster galaxies down to R=24. Among
these 15$\%$, about 1/4th are bulge galaxies and 3/4 are disk galaxies.
However, this contribution is spread over the whole field of view and
will act as a homogeneous background contribution of galaxies. It will
therefore not modify the relative variation of the estimated multicolor
types inside the Coma cluster. We could argue that dense regions of
filaments (without being real massive structures such as clusters or groups)
at z$\leq$0.2 could provide different galaxy type counts across the
field of view because they probably contain more early type
galaxies. However, even considering the largest known cosmic bubble sizes, at
least a dozen of such bubbles are superimposed between Coma and
z=0.2. The sum of all these bubbles therefore homogenizes the distribution
of field galaxy morphological types at z$\leq$0.2.

Similarly, we estimated that the contribution of galaxies in loose
groups represents $\sim$5$\%$ of the Coma cluster galaxies down to R=24
over the whole field of view. However, these contributions are
concentrated in precise locations, so locally, the contribution can be
much higher. Assuming the group GLF of Zandivarez et al. (2006) and the
galaxy spectral type estimates in low mass groups
($\leq$10$^{13.5}$M$_{\bigodot}$) of Dominguez et al. (2002), we
estimate that along the lines of sight to loose groups, galaxies in
groups can represent more than 85$\%$ of all z$\leq$0.2 galaxies. This
contribution is of the order of 50$\%$ for early types galaxies and can
reach 100$\%$ for other galaxy types.  As we cannot estimate precisely
the galaxy type contributions for the 5 loose groups detected behind
Coma, this clearly suggests that we have to remove these areas.

Maps of galaxy multicolor type density were then generated from a
simple count in cell technique, avoiding the masked regions and considering the
whole sample of galaxies down R=24. The
original cell size was 1' and we applied an additionnal smoothing of
3$\times$3 pixels in Fig.~\ref{fig:denspour}. We only took into account cells
  completely included in our field of view, so there are no edge effects.
This figure shows that
early type galaxies are spread over the whole field of view with the
possible exception of the north-west regions. Early and late Spirals
show localised density enhancements in the cluster outskirts. We note
however that the late (and perhaps early) spiral concentration located
north of the cluster center could be well explained by the loose G4
group. Later types (irregulars and starbursts) are also distributed
preferentially in the cluster outskirts. In particular, starburst
galaxies are mainly located at the northwest, where X-ray substructures
are also detected as an infalling group (see Adami et al. 2005a) and
where a concentration of early multicolor types is also detected. This
is also the place where the u* GLF is the steepest of the field (see
previous section).

The first thing to note in Fig.~\ref{fig:denspour} is that we confirm
that the Coma cluster is mainly populated by early type galaxies
($\sim$E+S0), as expected for such a rich structure: about 80$\%$ of the
galaxy population is made of ellipticals. Spirals represent only 15$\%$
of the faint galaxies.

Second, as expected and observed for bright galaxies (e.g. Whitmore et
al. 1993), there are later type galaxy clumps in the cluster
outskirts. This behavior essentially known for bright galaxies
(R$\leq$$\sim$18 for the Coma cluster) is now extended 6 magnitudes
deeper, and for a finer and more objective type separation.

Third, we also see a clear concentration of starburst objects in a place
where early spiral multicolor types are also detected and where
X-ray substructures are present. These starburst galaxies 
could see their star formation rate increased by interaction with the intra
cluster medium.

\section{Conclusions}

We have shown that we can reproduce the GLFs computed by Adami et
al. (2007a) by selecting cluster galaxies with PDFs from photometric redshift
techniques. This both puts the Adami et al. (2007a) conclusions on a
firmer ground, in terms of infalling directions and processes acting
on galaxies inside the cluster, and proves that the photometric redshift
technique (with a judicious cut in redshift) can help to select 
cluster members in nearby structures.

We clearly show that the u* GLF is steeper in the cluster
outskirts, as already suggested by Donas et al. (1995). This could be
due to a short burst of star formation in the faint galaxies, induced
by the pressure of the intracluster medium (see also e.g. Cortese at
al. 2007 or Bou\'e et al. 2008). 

We plan in a future paper to investigate how GLF and general galaxy properties
(based on photometric redshift estimates) vary with the local galaxy
density as derived for bright galaxies by Dressler et al. (1980).

\begin{acknowledgements}
The authors thank the referee for many useful and constructive comments.\\
We are grateful to the CFHT and Terapix (for the use of QFITS, SCAMP
and SWARP) teams, and to
the French CNRS/PNG for financial support. MPU
also acknowledges support from NASA Illinois space grant
NGT5-40073 and from Northwestern University. The authors thank 
J. Secker, P. C\^ot\'e and J.B. Oke for providing with us Keck spectra.\\
\end{acknowledgements}

\end{document}